\documentclass{svjour2}                    
\smartqed  
\usepackage{graphicx}
\newcommand{\ba}{\begin{eqnarray}}
\newcommand{\ea}{\end{eqnarray}}

\newcommand{\beq}{\begin{equation}}
\newcommand{\eeq}{\end{equation}}
\newcommand{\beqs}{\begin{eqnarray}}
\newcommand{\eeqs}{\end{eqnarray}}

\begin{document}

\title{ Gravitational Potential with Extra-dimensions
 and Spin Effects  In Hadronic Reactions   
}


\author{O.V. Selyugin       \and
        O.V. Teryaev 
}


\institute{Bogoliubov Laboratory of Theoretical Physics, \\
Joint Institute for Nuclear Research,
141980 Dubna, Moscow region, Russia \at
              Tel.: +07(496)-21-66481\\
              Fax: +07(496)-21-64668\\
              \email{selugin@theor.jinr.ru}           
}

\date{Received: date / Accepted: date}

\maketitle

\begin{abstract}
 The impact of the KK-modes in d-brane models of gravity
  with large compactification radii and TeV-scale quantum gravity
  on the hadronic potential at small impact parameters is examined.
  The effects of the  gravitational hadronic formfactors
  obtained from the hadronic generalized parton distributions (GPDs)
  on the behavior of the gravitational potential and
  the possible spin correlation effects  are  also analysed.
\end{abstract}

\section{Introduction}
\label{intro}

   Standard Regge representation shows that the energy dependence of the
   scattering amplitude depends on the spin of the exchanged $t$-channel particles.
   So the exchange of a graviton of spin  2 leads to a growth
   of the scattering amplitude with energy proportional to $s$ \cite{Hooft1}.
    As the usual gravitational interaction
   is very small, this growth can be seen at the Plank scale. However,
   a modern development of the fundamental theory, introduced long ago \cite{kk1,kk2},
   is connected with
 the fruitful idea \cite{Antoniadis:1990ew} that spacetime has a dimension higher than $D=4$
 manifested already at TeV scale \cite{Antoniadis:1998ig}.
 Now there are many different paths in the development of these ideas
 \cite{Mavr0}.
A number of studies of higher-dimensional (Kaluza-Klein) field theories were carried out
\cite{rev,rubakov1,perez,rand1}.
In a modern context,
the Kaluza-Klein (KK) theories arise naturally from (super)string theories
in the limit
where the relevant energies $E$ are much smaller than the string mass scale
$M_s \sim (\alpha')^{-1/2}$, $\alpha'$ being the slope parameter.
Since field theories of gravity
behave badly in the ultraviolet limit, Kaluza-Klein formulations should in general be
 regarded as effective actions, with an implicit or explicit ultraviolet
 cutoff $\Lambda$ \cite{han}.
  As a first approximation, we may suppose that all Standard Model fields
 are confined to a four-dimensional brane world-volume.
  In the Arkani-Hamed, Dimopoulos and Dvali approach (ADD) \cite{add}
 a large number $d$ of extra dimensions is responsible for a lower Planck
 scale, down to a TeV and only the graviton propagates in the $4+d$ dimensions.
 This propagation manifests itself in the standard $4$ dimensions as a tower of massive
 KK-modes. The effective coupling is obtained after summing over all the KK
 modes and, due to the high multiplicity of the KK modes, the effective interaction
has strength  $1/M_s$ \cite{nussinov,wells}.
Setting $M_{4+d}^{d+2}=(2\pi)^d \hat M_{4+d}^{d+2}$, as in
Ref. \cite{add} (motivated by toroidal compactification, in which the volume
of the compactified space is $V_d = (2\pi r_d)^d$) and applying Gauss's law at
$r<< r_d$ and $r >> r_d$, one finds that
$M_{Pl}^2 = r_d^d M_{4+d}^{2+d} \ \ \ $,
so that
$$  \ \ r_d =  \Biggl ( \frac{M_{Pl}}{M_{4+d}} \Biggr )^{2/d}/ M_{4+d} .$$
   In the higher-dimensional models with a warped extra dimension \cite{rand1},
  the first KK mode of the graviton can have a mass of the order of $1 \ $TeV
 and the coupling with matter on the visible brane
  is of the order $1 \ $TeV$^{-1}$.
    There are also "intermediate" models  with a small warp  which
  consider the brane  as almost flat. 
  Such models remove some cosmological bounds on the number of additional
  dimensions.
    All these models provide some experimental possibilities to check (or discover)
  the impact of the extra dimensions on our $4$-dimensional world.
  Now in many papers  new effects are examined which in principle can be seen
  at future colliders.

    In this work, we show that these effects can also be discovered in
  experiments on elastic polarized hadron scattering.
  We explore in this case the sensitivity of interference spin effects to
  small corrections to the scattering amplitude, linear rather than quadratic
  (as  in the case of cross sections) functions of a small parameter.
  We show that there will be some additional growth of the analyzing power
  $A_N$ which is connected with an additional term in the real part of the scattering
  amplitude, due to the graviton part of the hadron interaction.
  We will focus on $2$ extra dimensions, as different
  models give for that case coinciding results and as
  the TeV-scale mass does not contradict existing accelerator
  and cosmological bounds.

\section{The graviton contribution with KK-modes }
\label{sec:1}

Assuming that the higher-dimensional theory at short distances is
a string theory, one expects that the fundamental string scale $M_s$ and the Planck
mass $M_{4+d}$ are not too different (a perturbative expectation is that
$M_s \sim g_s M_{4+d}$).
   As of now, only known framework that allows a self-consistent description
   of quantum gravity is string theory \cite{polch-98}.
 Thus, a
compactification radius $r_d << M_{Pl}^{-1}$ corresponds to a
short-distance Planck scale and string mass $M_s$ which are  $<< M_{Pl}$.
So, we apply constrain on the quantum gravity scale $M_D$
  to the  string scale $M_s$.
There are different estimations of the mass $M_s$ from the particle
reactions and the cosmic data. The particle data lead to the low
bound of the mass $M_s > 0.5 \ $GeV,
 see for example \cite{Gullen,Perez-05}. The cosmic data gave significantly larger bound
 $M_s > 100 \ $ TeV \cite{Cosmic-data}. However, these estimations heavily
  depend on the model approaches. The large bounds were obtained at the tree level.
  If loop corrections are taken into account, the bound significantly decreases \cite{loop}.
  So, as in other works, we take the upper bound of the integrals proportional
  to $M_D = M_{4+d}$
   of an order of $1 \ $TeV.
Following \cite{nussinov,wells},
 the amplitude  taking into account the KK-modes can be written as
\begin{eqnarray}
   A_{grav.} &\sim&
\int_{0}^{M_D} \frac{d^{d-1} q_T}{ q^2 \ + \
q_T^2}  \nonumber \\
&=& \frac{M_D^d}{d} \frac{1}{q^2} {}_{2}F_{1} [1, d /2, 1 + d /2, -M_D^2 m^2/q^2])
\ea
  The hypergeometric function $_{2}F_{1}$ has a smooth behavior
  but the upper  limit of the integral appears  as a multiplicative coefficient
   $M_D^d$ which leads to a divergence of the Born amplitude
 if  $M_D^d \rightarrow \infty $ and $ 2 \leq d$.

\beq
\tilde A_{grav.}^{Born} =
\frac{\pi s^2}{M_{4+2}^4}\ln \Bigl ( 1 + \frac{M_D^2}{q^2} \Bigr )
\label{atildeneq2}
\eeq

\begin{figure}
\begin{flushleft}
 \includegraphics[width=0.5\textwidth]{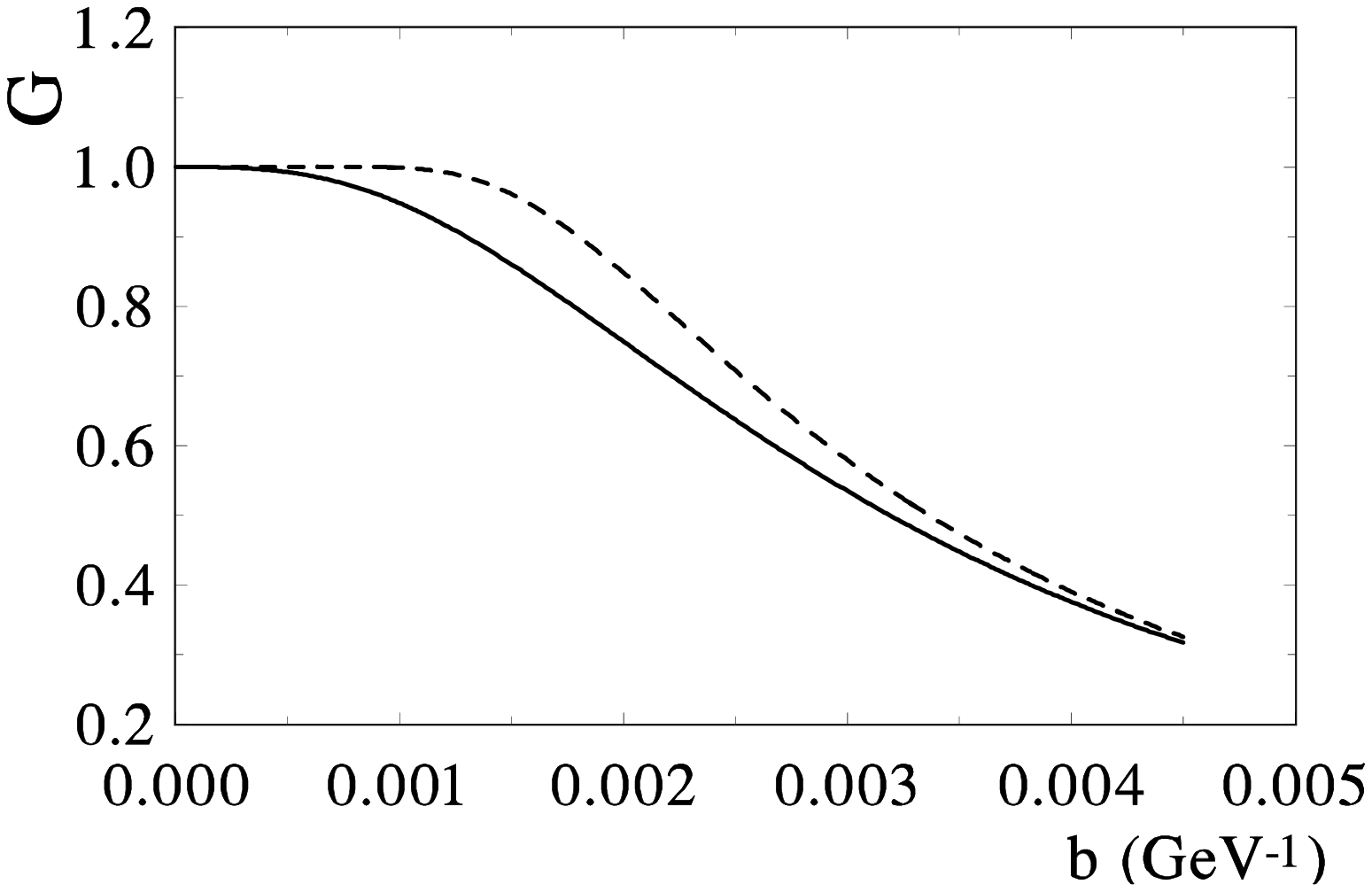}
\end{flushleft}

\begin{flushright}\vspace{-4.5cm}
 \includegraphics[width=0.5\textwidth]{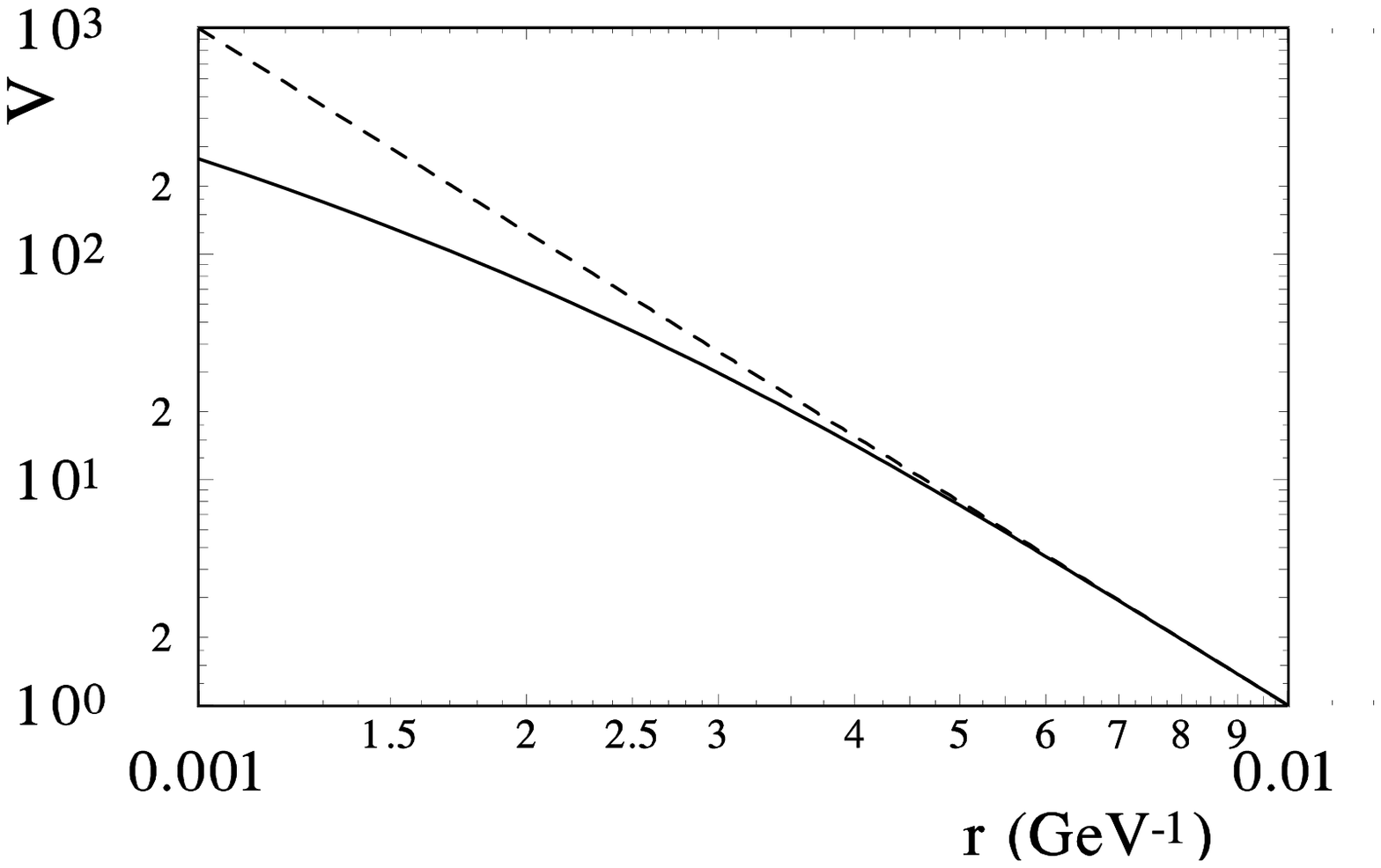}
\end{flushright}

\caption{a) [{\it left}]
  The profile function of proton-proton scattering  at  $\sqrt{s}=4 \ $TeV
  ({\it hard line} - (\ref{chi-our}) and {\it dashed line} - (\ref{chi-our})
  with correction term equal zero).
  b) [{\it right}] The graviton potential at small $r$ for $d=2$
 ({\it hard line} - our calculations, {\it dashed line}  $V(r) \sim 1/r^3$).
  }
\label{Fig_1}
\end{figure}

  In this work, we restrict our consideration only to the case $d=2$.
    If our particles live on the 3-dimensional brane,
   we can obtain the amplitude in an impact parameter representation,
  where the Born amplitude
  corresponds to the eikonal of the scattering amplitude \cite{Amati}
\begin{eqnarray}
  \chi(s,b) \ = \frac{1}{2 \pi} \int_{0}^{\infty} b \
  J_{0}(q b) \ A^{Born}(q^2) \ db.
  \label{born}
\end{eqnarray}
The potential of the two body interaction  corresponding to this eikonal
  can be obtained immediately:
\begin{eqnarray}
  V(r) = - \frac{2}{\pi r} \frac{d}{dr} \int_r^{\infty}
           \frac{b \ \chi(b)}{\sqrt{b^2 - r^2}} \ db.
\end{eqnarray}
 An exact calculation of (\ref{born}) for $d=2$ gives (see fig.1a)
\begin{eqnarray}
  \chi(b) = \frac{s}{2 M_{D}^4} \ (1 - b  M_D \  K_{1}(b \ M_D))/b^2.
\label{chi-our}
\end{eqnarray}
  Some detailed calculations can be found  in our work \cite{st-pr05}.
 It is interesting to note that when one takes the $KK$ modes (see fig. 1b) into
  account, one obtains the gravitational potential
\begin{eqnarray}
 V(r) \sim \frac{1}{r^3}(1-e^{-M_D  r} - M_D r e^{-M_D r})
\end{eqnarray}

\section{Nucleon-graviton interaction }

   First of all, we need to examine the gravitational interactions of point particles.
   There are several points of view on this.
   Some authors suppose that the interaction will be the same for different particles,
   and independent of their structure.
   At the same time, the equivalence principle \cite{KO} requires that the
   nucleon-graviton interactions
    are described by the matrix elements of energy momentum tensors,
    related to moments of Generalized Parton
   Distributions (GPDs) \cite{equiv1,equiv}.

   In  \cite{brod1}, the electromagnetic and gravitational form factors were calculated for
   the electron. It was shown that both  form factors have practically the same form.
   Of course, the proton has a complicated structure and there is no evidence for the
   equality of the electromagnetic and hadronic form factors. In some papers
    they have a different dependence on the momentum transferred to the hadron form factors.
   Based on the standard form of the parton distribution functions with a
   the specific  $t$ dependence, we obtained a good description of the electromagnetic
   form factors of the proton and neutron \cite{ST2a,ST2b}.
   On this basis we obtained the gravitational form factors $A(t)$ and $B(t)$ which
   correspond to GPDs $H(t,x)$ and $E(t,x)$.
   It can be shown that in the first approximation the  form factor $A(t)$ can be taken
    in the standard dipole form
   \begin{eqnarray}
   G(t) \   = \ 1 /( 1 - t /  \Lambda^2 )^2,
   \end{eqnarray}
    with $\Lambda^2 = 1.8 \ $GeV$^2$.
    In this case, our Born gravitational amplitude for $d=2$ will be
\beq
\tilde A_{grav.}^{Born} =
\frac{\pi s^2}{M_{D}^4}\ln \Bigl ( 1 + \frac{M_D^2}{q^2} \Bigr ) \ G^2(t)
\label{atildeneq2}
\eeq
   The corresponding gravitational potential is
\begin{eqnarray}
\tilde V_{grav.}^{Born}(r) \sim
\frac{1}{r^3} \Bigl ( 1 - (1+ \frac{\Lambda \ r}{20} \ (20+\Lambda \ r \ (9+\Lambda \ r))) e^{-\Lambda r} \Bigr ).
\label{vrd0}
\ea
  This leads to a change of the gravitational potential at small distances.
   For $d=0$, the gravitational  potential between two protons with
   taking into account the gravitational  form factor of the hadrons leads to the very different
   potential at distances of the order of the size of the hadron. This behavior has no divergence at
    $r=0$.

\begin{figure}
\begin{flushleft}
 \includegraphics[width=0.5\textwidth]{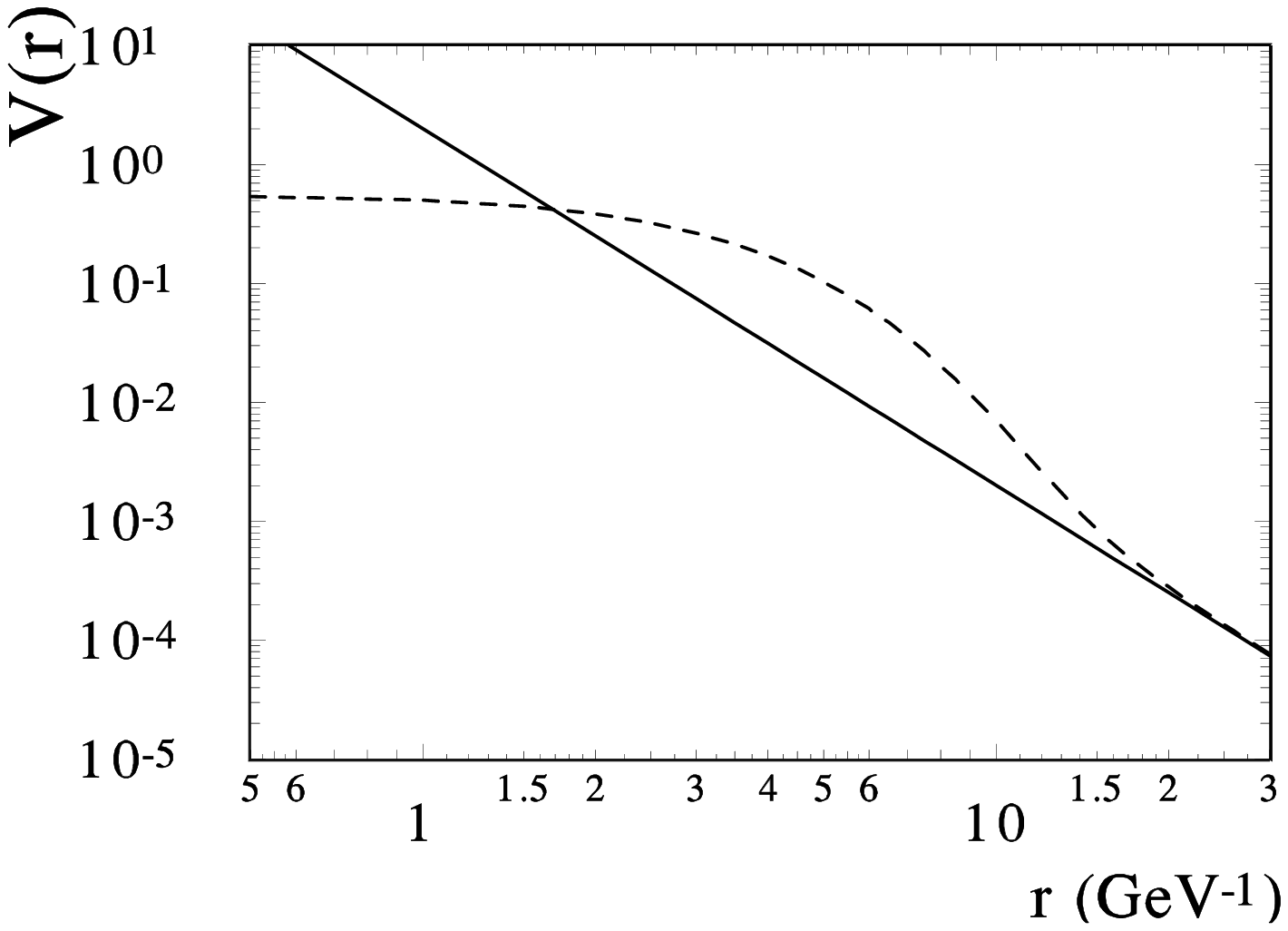}
\end{flushleft}

\begin{flushright}\vspace{-4.5cm}
  \includegraphics[width=0.5\textwidth]{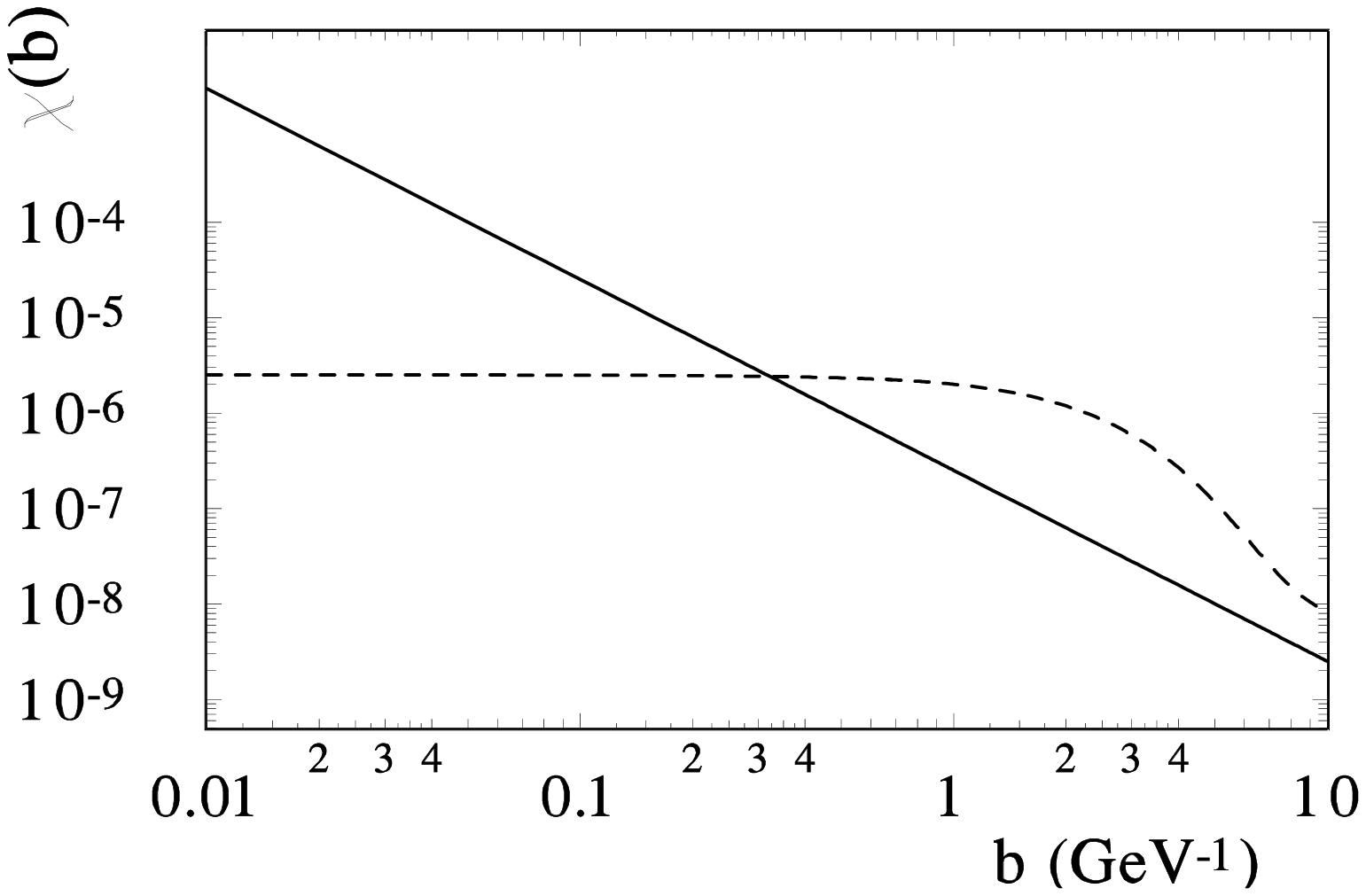}
\end{flushright}

\caption{a) [{\it left}]
  Gravi-potential without ({\it hard line}) and with ({\it dashed line}) taking into account the hadron
 form factor $A(t)$.
  b) [right]
  Behaviour of the eikonal
   over $b$  corresponds to the Born amplitude with $d=2$  ($1/b^2$ - {\it hard line};
     with gravitational hadron form factors - {\it dashed line}.
  }
\label{Fig_1}
\end{figure}

   For $d = 2$, the gravitational potential corresponding  to the Born amplitude
    is shown on Fig.2a. In this case, the form factors modify the amplitude at large distances,
    up to 20 times  the size of the proton.
    The decrease in the eikonal at small impact parameter is also large.
   This reflects the different behavior of the eikonal at small impact
  parameters.  The eikonal  for point particles  has the form  $1/b^2$ for $d=2$.
  The corresponding eikonal,
   including
   the hadron form factor
   but dropping the $1$ in
 the logarithm
  is
\begin{eqnarray}
 \chi_{grav}(b) & \sim & \int_{0}^{\infty} q \ J_{0}(qb) ln(\frac{M^2_{D}}{q^2}) G^2(t) \nonumber \\
 &=&\frac{\Lambda^2}{48} \Lambda^3 \ b^3 K_{3}(\Lambda \ b) ln(M_{D}^2) \ dq.
\label{eik2s}
\ea
  At large impact parameters it leads to the ordinary behavior  $b^{-2}$, but at small impact parameters
   this form removes the divergence of the standard case.
   The inclusion the $1$ in the logarithm leads to a complicated representation
   (Fig. 2b)
    which can be approximated by
\begin{eqnarray}
\tilde \chi_{grav}(b) = && \frac{s}{ M_D^4}
    \left[\frac{ ln(M_D^2) \Lambda^5  b^3}{1.3855*48} \ K_{3}(\Lambda b) 
        ( 1 + \frac{1 + 5 b^4}{2.2+10 b^4}) + \frac{0.5}{1+b^{1.5}} \right].
\label{eik2c}
\ea
Using this representation for the eikonal we can numerically  calculate  the full graviton amplitude
in the tree approximation and calculate the corresponding spin-correlation parameter
 $A_N$.

\section{ The spin correlation parameter $A_N$ }

The differential cross section and the analyzing power $A_N$  are
defined as follows:
\begin{eqnarray}
\frac{d\sigma}{dt}&=& \frac{2
\pi}{s^2}(|\Phi_1|^2+|\Phi_2|^2+|\Phi_3|^2
  +|\Phi_4|^2+4|\Phi_5|^2), \label{dsth}\\
  A_N\frac{d\sigma}{dt}&=& -\frac{4\pi}{s^2}
                 Im[(\Phi_1+\Phi_2+\Phi_3-\Phi_4) \Phi_5^{*})],  \label{anth}
\end{eqnarray}
in terms of the usual helicity amplitudes $\Phi_i$. These amplitudes
can be written as
\begin{eqnarray}
  \Phi_i(s,t) = \phi^h_{i}(s,t)
        + \phi_{i}^{\rm em}(t) \exp[i \alpha_{\rm em} \varphi_{\rm cn}(s,t)],
 \end{eqnarray}
where $\alpha_{\rm em}=1/137$ is the fine structure constant,
$\phi^h_{i}(s,t)$ describes the strong interaction of hadrons,
$\phi_{i}^{\rm em}(t)$  their electromagnetic interaction, and
$\varphi_{\rm cn}(s,t)$ is the electromagnetic-strong interference
phase factor.
This Coulomb-nuclear phase was calculated in the
entire diffraction domain taking into account  the form factors of
the  nucleons \cite{prd-sum}.
In this work, we define the hadronic and electromagnetic
spin-non-flip amplitudes as
%
\begin{eqnarray}
  F^{h}_{\rm nf}(s,t) = \frac{1}{2s}\left[\phi^h_{1}(s,t) + \phi^h_{3}(s,t)\right];  \
      F^{c}_{\rm nf}(s,t)
  = \frac{1}{2s}\left[\phi^{\rm em}_{1}(s,t) + \phi^{\rm em}_{3}(s,t)\right].
    \end{eqnarray}
 and the spin-flip amplitudes as
\begin{eqnarray}
  F^{h}_{\rm sf}(s,t) = \frac{1}{2s} \phi^h_{5}(s,t) ; \ \
    F^{c}_{\rm sf}(s,t) = \frac{1}{2s} \phi^{\rm em}_{5}(s,t).
    \end{eqnarray} Equation (\ref{anth}) was applied at high energy and at small angles.
 To take into account the unitarization effects, we used the standard eikonal
 representation for the spin-non-flip parts of the scattering amplitude  \cite{Sel-spin}.
 The  phase $\chi(s,b)$  is connected to  the interaction
 quasi-potential which can have real and imaginary parts
 \begin{eqnarray}
 \chi(s,b)\ = \ F_{\rm Born}(s,b)  \ \approx   \frac{1}{k}
   \ \int  \hat{V}\left( \sqrt{b^2  + z^2}  \right) dz.
 \label{potential3}
 \end{eqnarray}

  We have for the spin-non-flip
 \begin{eqnarray}
 F^{h}_{nf}(s,t) \ =  \  i
   \ \int_{0}^{\infty} \ b J_{0}(b q)\left[ 1- e^{\chi(s,b)}\right] \
  \ d b .
 \label{J1}
 \end{eqnarray}
  According to the standard opinion, the hadron spin-flip amplitude is
  connected with quark exchange between the scattering hadrons,
  and at large energy and small angles it can be neglected.
  Some models, which take into account non-perturbation effects,
  lead to a non-vanishing hadron spin-flip amplitude \cite{mog2a,mog2b,mog2c,mog2d}.
  Another complicated question is related to the difference between the
 phases of the spin-non-flip amplitude and of the spin-flip one.
  To estimate the possible graviton effect let us take
  the hadron-spin-flip amplitude to be proportional to
  the gravitational form factor $B(t)$ which is related to the GPD  $E(t,x)$
  \cite{ST1}:
\begin{eqnarray}
 F^{h}_{sf}(s,t) \ =  \  i K(s)B(t),
  \end{eqnarray}
where $K(s) =5.10^{-4}( ln(s) - i \pi/2$).

  Elastic nucleon scattering can occur  in the region of $t$
  after the second diffraction maximum  but
  still at
    small angles  $(t/s << 1)$.
    The analysis of the differential cross sections at  ISR
   energies has revealed scaling properties \cite{gazir}:
   the differential cross section can be described in a
   simple form     proportional to the electromagnetic form factor in the standard
   dipole form with $\Lambda = 0.71 \ $GeV which leads to the correct asymptotic
   behavior of the scattering amplitude  $\sim 1/t^4$.
   In this case, the eikonal will be represented as
   $\Lambda^5 b^3 K_{3}(b \Lambda)$.    The corresponding interaction constant is
   chosen to obtain the measured differential cross sections at $\sqrt{s} = 52.8 \ $GeV
   and $-t=10 \ $GeV$^2$.
 Taking into account an additional contribution from the graviton amplitude,
  we obtain for the spin-non-flip amplitude
 \beqs
\chi^{nf} (s,t) = \chi^{nf}_h (s,t) + \chi_{grav}(s,t).
\eeqs
  The additional contribution of the graviton  changes the real
  part of the spin-non-flip amplitude. Hence,   a difference appears
  between the spin-flip and the spin-non flip phases. So that  the
  analyzing power differs from zero (see Fig. 3).
   Of course, there is the question of the size of the mass cut $M_D$.
   The astronomical data lead to the maximum bound  $M_D \geq 1500 \ $TeV.
   However, the particle physics data lead to the possible size  $M_D \sim 1 \ $TeV
   (see the corresponding discussion \cite{Mavr1,wells1,perez}). So we will examine
   the possible polarization effects for  $M_D$ in the region of a few TeVs.
   Of course, larger $M_D$ require   interactions at larger energy.
\begin{figure}
\begin{flushleft}
  \includegraphics[width=0.5\textwidth]{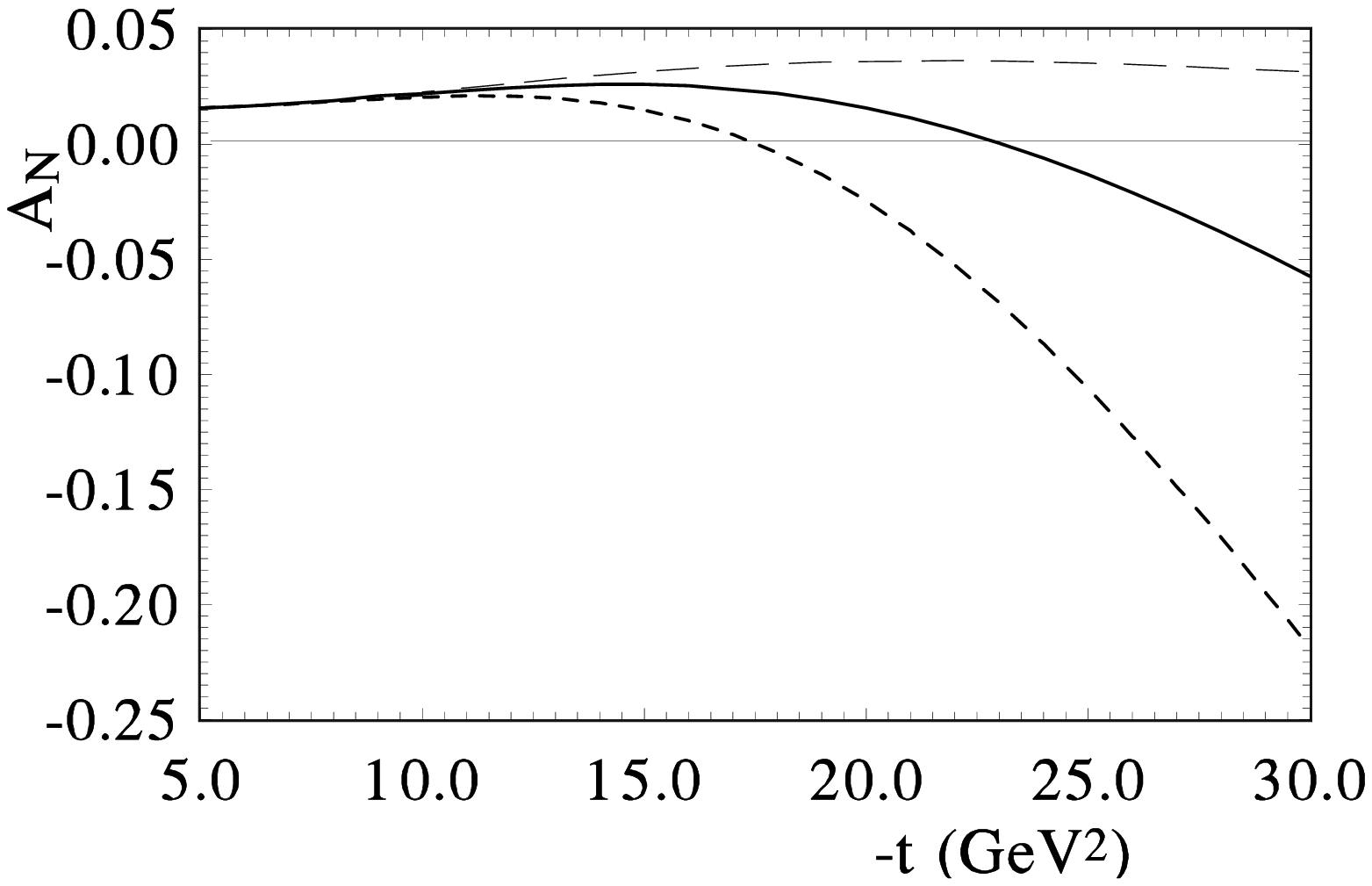}
\end{flushleft}

\begin{flushright}\vspace{-4.5cm}
  \includegraphics[width=0.5\textwidth]{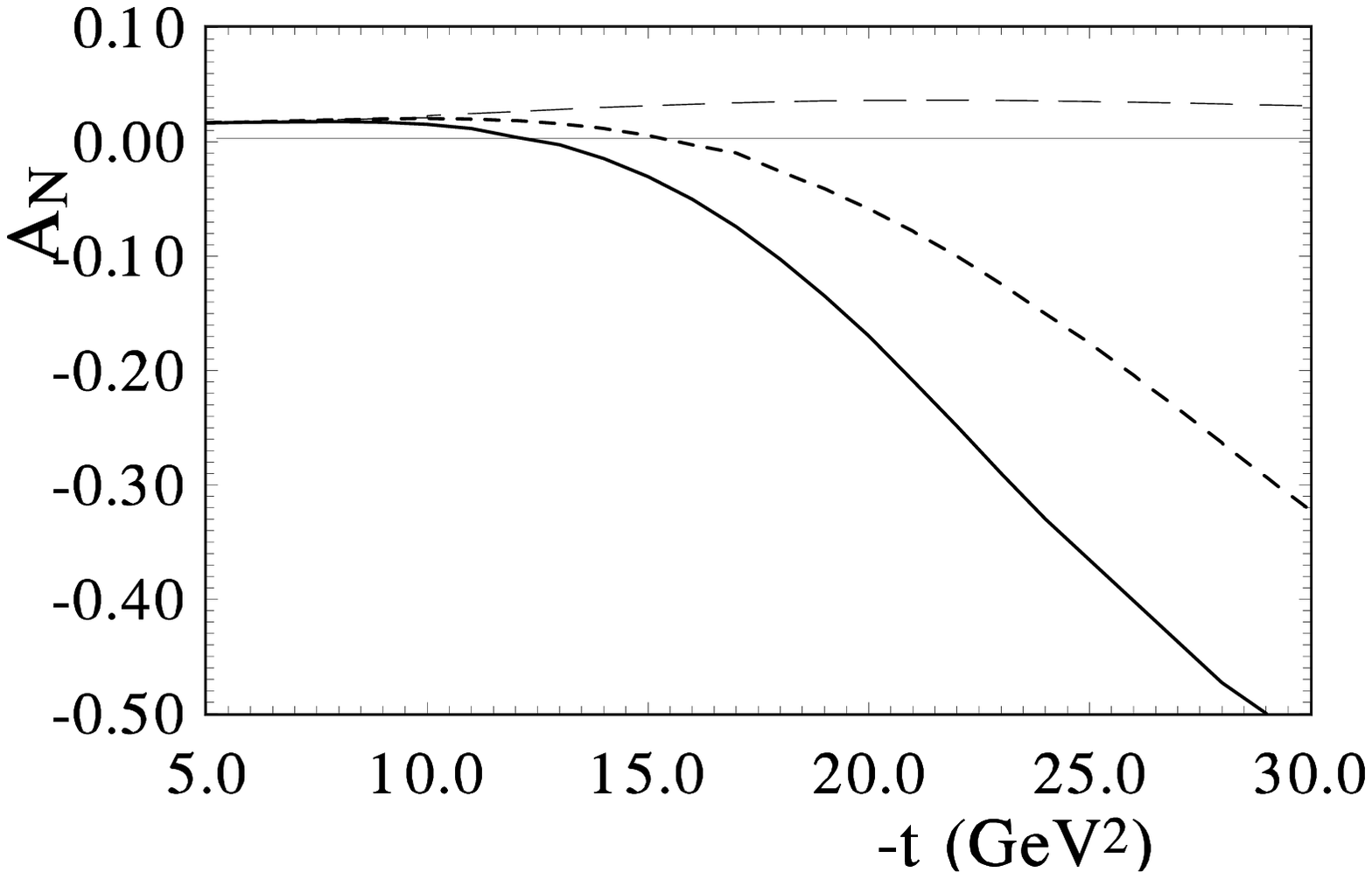}
\end{flushright}
\caption{
  $A_N$ calculated without taking into account the gravitational
  form factor $A(b)$:
 [left]
at $\sqrt{s} \ = \ 540 \ $\,GeV  (short dashed line ---
  without graviton amplitude;
 hard and long-dashed lines --- with $M_D=1.5 \ $TeV and $M_D=2 \ $TeV);
  [right]
   the same at  $\sqrt{s}  =  1.8  $\,TeV
   (hard and long-dashed lines --- with $M_D=2. \ $TeV and $M_D=2.5 \ $TeV).
  }
\label{Fig_1}
\end{figure}

  Figure 4 displays  $A_N$ calculated for  two values of
  $M_D$, $ 1.5$ TeV and $M_D = 2 \ $TeV.
  Despite the fact  that the measurement of
  the elastic cross section at such values of the momentum transfer is
  not a simple task, we can see that the size of the effect may be sufficient for a
  discovery. The dashed line in this figure shows the analyzing power
  without 
  gravitation interaction.
 A slight difference of this curve from
  zero is due to the small contribution of the electromagnetic amplitude.
          The main  characteristic of the analyzing power is its $t$ dependence.
  Such strong $t$ dependence 
  will be present in all cases.
   The form of $A^{gr}_N$
  is  dictated by the form of the graviton amplitude, and in the cases when
  the number of additional dimensions exceeds two ($ d > 2$), such a behavior will be   more pronounced.

\begin{figure}
\begin{center}
  \includegraphics[width=0.5\textwidth]{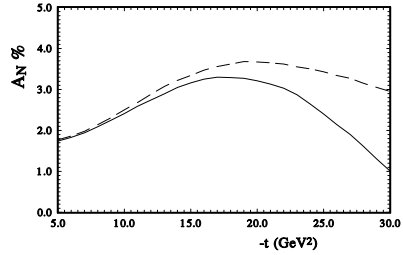}
\end{center}

%
\caption{
  $A_N$ calculations without and with  the gravitational  hadron form factors $A(b)$
  at $\sqrt{s}  =  4 $\,TeV
  (hard  lines - with $M_D=1. \ $TeV;
   dashed line -  without graviton amplitude).
  }
\label{Fig_1}
\end{figure}

\section{Conclusion}

     Our calculations of the impact of two additional dimensions  on the gravitational
     potential at small distances  show that it differs from  a simple power
      at $r \leq 3 \ $GeV$^{-1}$ and it changes
     the  profile function of  proton-proton scattering.
     Including  the effects of Kaluza-Klein modes of graviton scattering   amplitudes,
     with two extra dimensions and taking into account  the gravitational form factor,
     which we calculated from the GPDs of the nucleons,
     it was shown  that   the impact-parameter dependence of the gravitational eikonal
     strongly deviates from the standard $1/b^2$  dependence.
     This is the main result of our paper.
     We think that this effect has to be taken into account in the calculation of the production
     of Black Holes  at  super-high energy accelerators.

     We have  shown that the gravitational interaction  additional dimensions and
     the  possible small spin-flip amplitude, proportional to the gravitational
     form factor $B(t)$, lead to large spin correlation effects at small angles and
      $-t \sim 10-30 \ $GeV$^2$. However, the inclusion of  the gravitational form factors
      $A(b)$ decreases this effect  and drastically changes its form.

{\small This work was supported in part by Grants RFBR 09-01-12179
and RF MSE RNP 2.2.2.2.6546.

\end{document}